\documentclass[preprint,aps,prd,showpacs,nofootinbib,tightenlines]{revtex4}

\usepackage{amsmath} 

\def\spose#1{\hbox to 0pt{#1\hss}}
\def\ltapprox{\mathrel{\spose{\lower 3pt\hbox{$\mathchar"218$}}
 \raise 2.0pt\hbox{$\mathchar"13C$}}}
\def\gtapprox{\mathrel{\spose{\lower 3pt\hbox{$\mathchar"218$}}
 \raise 2.0pt\hbox{$\mathchar"13E$}}}
\def\inapprox{\mathrel{\spose{\lower 3pt\hbox{$\mathchar"218$}}
 \raise 2.0pt\hbox{$\mathchar"232$}}}
\def\eqn#1{\label{eq:#1}}
\def\Equation#1{Equation~(\ref{eq:#1})}

\def\eq#1{Eq.~(\ref{eq:#1})}

\def\chpt{\raise0.4ex\hbox{$\chi$}PT}
\def\schpt{S\raise0.4ex\hbox{$\chi$}PT}

\def\cd{{\cal D}}

\def\psibar{\overline\psi}

\begin{document}

\title{Comment on ``Chiral anomalies and rooted staggered fermions''}

\author{Claude Bernard}
\email[]{cb@lump.wustl.edu}
\affiliation{%
Department of Physics, Washington University,
St. Louis, MO 63130, USA}

\author{Maarten Golterman}
\email[]{maarten@stars.sfsu.edu}
\altaffiliation[Permanent address: ]{\baselineskip 12pt Department of Physics and Astronomy,
San Francisco State University,
San Francisco, CA 94132, USA}
\affiliation{%
Grup de F{\'\i}sica Te{\`o}rica and IFAE,
Universitat Aut{\`o}noma de Barcelona, 08193 Barcelona, Spain}

\author{Yigal Shamir}
\email[]{shamir@post.tau.ac.il}
\affiliation{%
School of Physics and Astronomy,
Raymond and Beverly Sackler Faculty of Exact Sciences,
Tel-Aviv University, Ramat~Aviv,~69978~Israel}

\author{Stephen R. Sharpe}
\email[]{sharpe@phys.washington.edu}
\affiliation{%
Physics Department, University of Washington,
Seattle, WA 98195-1560, USA}

\date{\today}

\begin{abstract}
In hep-lat/0701018, Creutz claims that the rooting
trick used in simulations of
staggered fermions to reduce the number of tastes misses key physics
whenever the desired theory has an odd number of continuum flavors, and uses this
argument to call into question the rooting trick in general.
Here we show that his argument fails as the continuum limit is approached, and
therefore does not imply any problem for staggered simulations.
We also show that the cancellations necessary to
restore unitarity in physical correlators in the continuum limit
are a straightforward consequence of the
restoration of taste symmetry.

\end{abstract}


\maketitle

\noindent {\bf 1.}
Simulations of QCD using staggered fermions have, in the last few years,
attained impressive precision
 in calculations of the hadron spectrum
and matrix elements~\cite{davies,MILCmq,MILCfKpi,MILCsemil,MILCfD}.
A crucial part of the methodology is the need to use the so-called
``fourth-root trick.'' Each flavor of staggered fermion on the lattice represents,
in the continuum limit, four degenerate continuum fermions (conventionally
called ``tastes''). To reduce to the desired number of continuum fermions---one
per flavor---the simulations are done with the fourth root of the
(four-taste) staggered fermion determinant. Since the staggered Dirac operator
does not represent four degenerate fermions for $a\ne 0$ ($a$ is the lattice
spacing), the validity of this procedure is a non-trivial issue.
In the past year, there has been significant progress on this issue,
showing both the nature of the discretization effects at non-zero lattice spacing
and how, given certain technical assumptions,
the correct physics is obtained in the continuum 
limit~\cite{BGS,shamirRG,CB,Sharpe-review,BGS-lat06}.
The required assumptions are plausible (at least to us), but
have not yet been proven analytically or tested in detail numerically. 

This note concerns particular criticisms of
the rooting procedure given by Creutz \cite{creutz07}.
He argues that the rooting procedure ``mutilates the expected continuum
holomorphic behavior in the quark masses'' and is manifestly flawed when applied
to a theory with an odd number of flavors.
Given the phenomenological significance of the results obtained with
staggered fermions, it is important to determine the correctness and applicability
of his arguments.  
Here we show that his arguments do not apply to
rooted staggered fermions as they are used in simulations of QCD.
In other words, Creutz's criticisms do not call into question
the correctness of the physical results obtained in the continuum limit
of staggered QCD with the rooting procedure. 

\bigskip
Before going into details, we stress several points:

\begin{itemize}

\item[A.\ ]{}  
Simple, well-tested
properties of the staggered theory are all that are required to 
refute Ref.~\cite{creutz07}.

\item[B.\ ]{} 
Showing that Creutz's reasoning is incorrect or misleading does not of course prove
the correctness of the rooting procedure, 
but only that his particular criticisms do not apply.\footnote{%
We note that Creutz's work~\cite{creutz07} 
does not respond to the arguments of Refs.~\cite{BGS,shamirRG,CB,Sharpe-review,BGS-lat06}
supporting the validity of rooting,
and indeed does not refer to several of these references.}

\item[C.\ ]{} We do not dispute that the structure of the rooted staggered theory is
complicated (``ugly'' \cite{Sharpe-review}) at non-zero lattice spacing, $a$.
As we have emphasized \cite{BGS,shamirRG,CB,Sharpe-review,BGS-lat06},
there are certainly violations of unitarity and locality for $a\ne0$.  
However, all that is required for
existing and planned staggered simulations to compute QCD quantities correctly
is that the rooted staggered theory reproduces QCD {\it in the continuum limit}\/ for fixed,
positive 
quark masses.
Any ``mutilations'' or problems listed by Creutz at non-zero
$a$ that go away in the continuum limit are not related to the issue of correctness.
Indeed, such discretization effects would be relevant at all only to the extent they
were not understood and therefore not properly taken into account in final systematic
error estimates after extrapolation to the continuum.  As explained below,
that is not the case.

\item[D.\ ]{}  Several properties 
Creutz finds ``peculiar,'' ``contrived,'' or ``absurd,'' such as the lack of commutativity of
an ultraviolet and an infrared limit, or the existence of unitarity violations
outside of the physical subspace, are not special to the 
rooted staggered regulator, but can appear with other regulators or in 
manifestly correct continuum versions of QCD. We give examples below.

\end{itemize}

\medskip
\noindent {\bf 2.}
Creutz's central argument can be summarized as follows.
QCD with an odd number of degenerate quarks is not invariant under a change
in the sign of the (common) quark mass, $m$~\cite{CP}.
The corresponding rooted staggered theory, by contrast, is invariant
under this change at any lattice spacing
(since it descends from a theory with four tastes per flavor, which is
itself invariant). Thus, taking the continuum limit, the lattice theory cannot
correctly represent the desired underlying theory for {\em any}\/ $m$.
In particular, odd powers of $m$ are missing in the chiral expansion.
Creutz claims that this argument is particularly sharp at finite volume,
because one then avoids the discontinuities that can develop in infinite
volume. In finite volume, one expects in the continuum limit that
physical quantities for theories with an odd number
of quarks will be analytic functions of $m$ having
both even and odd powers.
Such functions cannot, Creutz argues, be obtained
from the rooted lattice theory whose partition function
is expandable in (only) even powers of $m$
at any lattice spacing.

Two key points are needed to understand the flaws in this argument.
The first is that rooted staggered fermions
correspond, in the continuum limit, to a theory with a positive
quark mass, irrespective of the sign of the mass in the
underlying four-taste quark determinant.
Schematically,
\begin{equation}
\left[{\rm det}(D_{\rm stag}+m)\right]^{N_f/4} \
\xrightarrow[\ a \rightarrow 0\ ]{}\
\left[{\rm det}(D_{\rm cont} + |m|)\right]^{N_f}
\,,
\label{eq:|m|}
\end{equation}
where $D_{\rm cont}$ is a continuum Dirac
operator for one quark field, and $N_f$ is the number of flavors.
This equation is schematic
as we are not
being careful with overall factors or mass renormalization, neither
of which are relevant here as they do not change the symmetry
properties under a change in sign of $m$.
However, for those who might worry that problems could lurk in this
schematic form, we note that it is possible to define precisely 
the form in the continuum limit using renormalization group techniques ---
see Ref.~\cite{shamirRG}.
Note that the root on the left-hand side of
\eq{|m|} is always chosen to be positive.\footnote{%
Creutz states that ``attempts to define the rooting
procedure by selecting one fourth of the eigenvectors will
necessarily involve ambiguities.'' We agree with this statement,
but stress that this is {\em not} what is done in the rooting trick,
and thus is not relevant to the issue of its correctness.
Rooting is an unambiguous procedure that takes the unique positive
fourth root of ${\rm det}(D_{\rm stag}+m)$, which is strictly positive for $m$ real and
non-zero. The effect of rooting is to replace each approximate quartet of
eigenvalues by its geometric average.
Note that when a sufficiently large chemical potential, $\mu$, is turned
on, one must deal with the (now complex) individual eigenvalues
of $D_{\rm stag}$, and as a result ambiguities do arise when taking the fourth
root of the determinant~\cite{GSS}. However, this does not apply in the case of
$\mu=0$ considered here.}
\Equation{|m|} implies that the physical
theory with negative quark masses is not directly accessible with the
rooting trick.

The second basic point is that rooted staggered fermion determinants are,
by construction (because of the root),
non-analytic functions of the quark mass, {\em even in finite volume}.\/
In particular, in the continuum limit, they are non-analytic at
$m=0$.

We elaborate on these points below, but first show how they refute
Creutz's argument. This can be seen from \eq{|m|}.
In a rooted staggered simulation, the left-hand side
(which is the fermion contribution to the weight for a given
gauge configuration) is indeed even in $m$, as in Creutz's argument.
But the proposed continuum limit, given by the right-hand side,
is also even. Thus there is no contradiction.
There is no problem with the right-hand side being non-analytic in $m$ because
the left-hand side is non-analytic by construction. In particular, in finite volume,
physical observables
in the rooted staggered theory can have both even and odd powers of
$|m|$ in the continuum limit.

This argument shows that the rooting procedure {\em can} lead
to the desired continuum dependence on $m$ (for $m>0$) when $a\to 0$.
There is no inconsistency.
Whether it {\em does} lead to the desired dependence is 
a different issue. This requires that taste symmetry be
restored in the continuum limit, for which there is, by now,
extensive evidence.
For example, Fig. 1 of Ref.~\cite{MILC-Lat06} shows not only that the splitting
of various-taste pions goes to zero as the continuum limit is approached, but also
that the rate of approach to zero (as $\alpha^2_S(a)a^2$)
is in agreement with  the theoretical prediction for the version of the staggered action used.
The restoration of taste symmetry is also found directly in
numerical studies of the eigenvalues, which fall into approximate quartets
for eigenvalues well below the cutoff, with the splittings decreasing toward zero 
as $a$ is reduced~\cite{NUMERICAL-EIGENVALUES1,NUMERICAL-EIGENVALUES2}.
Further, these studies show that, in topologically non-trivial configurations,
what would be zero modes in the continuum (or with an overlap lattice
operator) appear as quartets of near-zero staggered eigenvalues.
Note that there are also good theoretical reasons for believing that
taste symmetry is restored in the continuum limit.  In particular, it has long been
known~\cite{KLUBERG,GoltSmit}  
that 
taste violations are due to an operator with dimension greater
than four.\footnote{%
The argument has been made more
rigorous in Refs.~\cite{shamirRG,Sharpe-review}.  Recent calculations
have also shown that the size of taste violations can be understood quite well
perturbatively~\cite{LEPAGE}.}

We now illustrate how the expected behavior in the continuum limit
is approached, assuming the restoration of taste symmetry.
As stressed by Creutz, the dependence on the sign of $m$
in the continuum arises from zero modes of $D_{\rm cont}$.
Consider a gauge field with winding number $\nu=1$,
so that the continuum determinant can be formally written as
\begin{equation}
\left[{\rm det}(D_{\rm cont} + m)\right]^{N_f}
= \left[m\, F_{\rm cont}(m^2)\right]^{N_f}
\,,
\end{equation}
where the factor of $m$ comes from the zero mode of $D_{\rm cont}$, and
$ F_{\rm cont}(m^2)$ is the contribution of the non-zero eigenvalues.
Because the latter come in imaginary pairs $\pm i \lambda$, $ F_{\rm cont}(m^2)$
is necessarily a function of $m^2$  and is strictly positive for $m\not=0$.
We are imagining working in finite volume
so that the spectrum is discrete.
For odd $N_f$ the contribution of such configurations
is manifestly odd in $m$.

How does the rooted staggered theory approach this limit?
Like the non-zero eigenvalues of $D_{\rm cont}$, the eigenvalues of
$D_{\rm stag}$ are imaginary and occur in complex conjugate pairs.
The low-lying eigenvalues
form into quartets, and these quartets
become degenerate in the continuum limit. 
For a configuration with winding number unity there thus will be
two pairs of eigenmodes with eigenvalues of $O(a)$,
corresponding to one zero mode per taste in the continuum limit. 
Labeling these as
$\pm ia\lambda_{1,2}^{\rm (stag)}$,
the rooted staggered determinant is
\begin{equation}\label{eq:lattice-root}
\left[{\rm det}(D_{\rm stag}+m)\right]^{N_f/4}
=
\left[\{(a\lambda_1^{\rm (stag)})^2 + m^2\}
\{(a\lambda_2^{\rm (stag)})^2 + m^2\}\, F_{\rm stag}(m^2)
\right]^{N_f/4}
\,.
\end{equation}
Like $ F_{\rm cont}(m^2)$,
$F_{\rm stag}(m^2)$ is strictly positive for $m\not=0$.
As claimed by Creutz, \eq{lattice-root}
is manifestly an even function of $m$.
If we stay in a finite volume,
the determinant is analytic in the complex $m$ plane aside from cuts along the
imaginary-$m$ axis. In particular, its expansion about $m=0$ contains
only even powers of $m$, for $a>0$.
If we take the continuum limit,\footnote{%
Which is to be done at fixed physical volume.}
however, the nearest cuts
(due to the near-zero eigenvalues $\pm i a\lambda_{1,2}^{\rm (stag)}$)
move to the origin, and, focusing on the zero-mode sector,
one finds
\begin{equation}
\left[{\rm det}(D_{\rm stag}+m)\right]^{N_f/4}
\left[F_{\rm stag}(m^2)\right]^{-N_f/4} \
\xrightarrow[\ a \rightarrow 0\ ]{}\
|m|^{N_f}
\,.
\end{equation}
The key point is that the non-analyticity
intrinsic to a root leads to a positive contribution from the
zero modes, irrespective of the sign of $m$. In particular, we
see how, for odd $N_f$, odd powers of $|m|$ arise in the continuum limit due
to this non-analyticity.

That the rooting trick leads to a continuum theory with a positive
mass is not related to the use of staggered fermions.
Consider the lattice theory obtained by taking the positive
$N_f/4$ power of the determinant for four degenerate ``tastes''
of overlap fermions. There are exact zero modes
in this theory, and the contribution of each to the determinant
is
\begin{equation}
(m^4)^{N_f/4} = |m|^{N_f}
\,.
\end{equation}
The remaining eigenmodes lead to an even function of $m$.
For positive $m$, the rooting has no impact, and one obtains
the correct continuum limit, with, for odd $N_f$, both even and odd
powers of $m$. For negative $m$, one obtains the continuum theory
with positive mass $|m|$. All this holds in any finite volume.
Thus the rooted overlap theory provides a clear
example of the failure of Creutz's argument.

Comparing the rooted staggered and overlap theories, one sees that
to reproduce the form of the overlap result, it is essential
to take the continuum ($a\to 0$) limit before the chiral ($m\to0$)
limit. This fact is already well known in the 
literature~\cite{NUMERICAL-EIGENVALUES1,SMIT-VINK,LIMITS}.
Indeed, with or without the fourth root, staggered fermions
do not always produce correct continuum physics if one takes
the $m\to0$ limit before the $a\to0$ limit.%
\footnote{Note that, while Ref.~\cite{LIMITS} argues
that the continuum and chiral limits commute for many common observables,
other observables for which these limits do not commute are also discussed.}
Thus the problems that Creutz finds in the staggered theory when a quark mass
is set to zero before taking the continuum limit are irrelevant to simulations,
for example those of the MILC collaboration \cite{davies,MILCmq,MILCfKpi,MILCsemil,MILCfD},
in which the quark masses are kept positive for all lattice spacings, and any extrapolations
are performed in the proper order.

We note that the effective low energy theory for
(rooted or unrooted) staggered quarks,
staggered chiral perturbation theory \cite{SCHPT},
provides an explicit framework to see both that the correct
continuum physics is obtained, and that the continuum
theory is always that with positive mass.
The chiral theory for rooted staggered quarks, including
discretization effects,
is obtained using a limiting procedure with an even
number of flavors $\!\times\!$ tastes at each step.
This necessarily results in a theory which is invariant under
a change in sign of $m$: for $m<0$ one can perform a non-singlet
axial rotation
to make $m$ positive since there is
an even number of flavors $\!\times\!$ tastes.
Nevertheless, the effective theory in the rooted case
goes over as $a\to 0$ to the correct, positive quark mass,
continuum chiral theory,
for both odd and even numbers of flavors.\footnote{%
At the moment, the effective theory calculations can only be traced from negative to
positive masses in infinite volume. In finite volume, for small enough
masses one necessarily enters the epsilon-regime ($m_\pi L \ll 1$), for which
a non-perturbative calculation is needed in the effective theory. In that case,
a non-perturbative generalization of the replica trick is needed to
incorporate rooting into the effective theory; development of such a generalization
is still in progress \cite{BGS-lat06}.}
For example, the masses of pions at leading order take the form
$m_\pi^2 = 2 B_0|m| + O(a^2)$, with $B_0$ the continuum low-energy constant.
The continuum theory with negative mass (which, for $N_f=3$, displays
spontaneous CP violation~\cite{CP}) is not directly accessible in the chiral
theory for rooted staggered quarks,
just as at
the quark level.\footnote{%
The odd-$N_f$, continuum negative-mass theory can be reached,
at least in principle, by adding an explicit theta term with $\theta=\pi$ to the rooted
staggered action. While numerically difficult in QCD due to the sign problem,
this approach has been shown to work beautifully in
the Schwinger model \cite{DURR}.}

\medskip
\noindent {\bf 3.}
Another feature of the rooted staggered theory that bothers Creutz is the existence of
multiple versions of the physical pions, with various tastes.  Indeed, due to the exact
non-singlet chiral symmetry of staggered fermions in the massless limit, one pion per flavor
is a true (pseudo-)Goldstone boson, even in situations where no Goldstone particle should
exist:  a one-flavor theory, for example.   
While Creutz recognizes that this feature may
be repaired in the continuum limit by cancellations from pions of other tastes, he 
finds the suggested mechanism of repair an ``unproven conjecture''  that is
``rather contrived.''  
On the contrary,
the mechanism depends {\it only}\/ on the restoration of taste symmetry in the continuum,
for which there is a great deal of evidence, as discussed above.  To see how taste symmetry
fixes this, and many similar ``problems,'' it is necessary simply to examine the 
continuum-limit theory.\footnote{%
The ``reweighted theories''~\cite{shamirRG} provide a way to
give this limit a rigorous meaning. The result
is a formulation with a more complicated gauge sector than in
the following discussion, but this difference does not affect the
fermionic symmetries being considered.
See also the discussion of ``valence rooting''
in Ref.~\cite{Sharpe-review}.}
Consider the Euclidean QCD partition function in a formal
continuum setup,
\begin{eqnarray}
  Z_{cont} &=& \int \cd A\;\cd\psi\;\cd\psibar\;
  \exp\Big(-S_g + \sum_{i=1}^{N_f} \psibar_i (D+m_i) \psi_i \Big)
\label{eq:contint}
\\
  &=& \int \cd A\; \exp(-S_g)\; \prod_{i=1}^{N_f} \det(D+m_i)\,.
\label{eq:cont}
\end{eqnarray}
Here $S_g(A)$ is the gauge action,
$D$ is the massless Dirac operator,
and  $m_i$ is the mass of the $i^{\rm th}$ flavor. We assume all $m_i$
are positive, which implies that the (formal) continuum determinant,
$\det(D+m_i)$, is positive.
Now endow each quark flavor with a new degree of freedom, called taste,
ranging from 1 to $N_t$, and replace the Dirac operator $D+m_i$
by $(D+m_i) \otimes {\bf 1}$, where ${\bf 1}$ is the $N_t \times N_t$
identity matrix.\footnote{%
For the continuum limit of staggered fermions, $N_t=4$,
but the discussion in the continuum holds more generally.}
Since, evidently,
\begin{equation}
  {\rm det}^{1/N_t}\Big((D+m_i)\otimes {\bf 1} \Big)
  = \det(D+m_i) \,,
\label{anl}
\end{equation}
where we take the positive $N_t^{\rm th}$ root,
it follows that
\begin{equation}
  Z_{cont} = \int \cd A\; \exp(-S_g)\;
  \prod_{i=1}^{N_f} {\rm det}^{1/N_t}\Big((D+m_i)\otimes {\bf 1} \Big)\,.
\label{eq:rootNt}
\end{equation}
Clearly, the combined effect of the two modifications is that
nothing at all has changed.  The $N_t^{\rm th}$-root theory
derived from the $N_t$-taste quark fields is nothing but the original
theory in disguise.
 Yet, the extra redundancy
provided by the new taste index provides a source for apparent paradoxes.

The paradoxes all have their origin in the following fact. The theory
with $N_f N_t$ quarks, whose $N_t^{\rm th}$ root is being taken,
has, nominally, a much bigger symmetry group than that of the original theory.
To see this explicitly one can add a (meson)
source to the fermion determinant~\cite{CB}, giving
\begin{eqnarray}
\lefteqn{  {\rm det}^{1/N_t}\Big([(D+M)\otimes {\bf 1}] + J \Big)\; =}
\nonumber\\
&=&
  {\rm det}^{1/N_t}[(D+M)\otimes {\bf 1}]\;
  \exp\bigg[ \frac{1}{N_t}\, {\rm tr} \log \Big(
  1 +  J [(D+M)\otimes {\bf 1}]^{-1} \Big) \bigg]
\label{eq:detJI}\\
&=&
  \det(D+M)\;
\bigg[ 1 + \frac{1}{N_t} \, {\rm tr}\Big( J [(D+M)\otimes {\bf 1}]^{-1} \Big) 
+ \dots \bigg]
\,.\label{eq:detJII}
\end{eqnarray}
Here the massless Dirac operator $D$ and the mass matrix
$M={\rm diag}(m_1,m_2,\ldots,m_{N_f})$
carry spin, color and flavor indices but no taste index.
The source is a color singlet, and has a general spin,
flavor and taste structure.
Because of the taste structure of the source,
taking the $N_t^{\rm th}$ root is not entirely trivial.
But since the source is used solely to generate correlation functions,
we are always allowed to expand as shown.
Thus, as before, we only need the $N_t^{\rm th}$ root
of the determinant of the taste-invariant operator $(D+M)\otimes {\bf 1}$,
which is analytic and positive.

The net result, \eq{detJII}, shows that one obtains
correlation functions of propagators evaluated with the correct
QCD determinant.
One can now derive Ward identities in the enlarged flavor-taste space
by doing transformations on $J$. For example,
if all masses are non-zero and no two masses are equal,
there is a $U(N_t)$ taste symmetry for each flavor, leading to
$N_f N_t^2$ conserved vector charges.
This should be compared to the $N_f$ conserved charges in the original theory.
If the mass of one particular flavor goes to zero, e.g.\ $m_u\to 0$,
then the up-sector taste symmetry enlarges to
an $SU(N_t)_R\times SU(N_t)_L\times U(1)_V$
non-anomalous chiral symmetry, while in the original theory
there is still only one conserved charge in the up sector.
The extra symmetries of the rooted theory mean that there are many extra
Goldstone bosons (pseudo-Goldstone bosons in the massive case) compared to
the original theory.  This seems paradoxical, since the rooted
version, \eq{rootNt}, was supposed to be completely equivalent to
the original theory, \eq{cont}.

However, the extra Goldstone bosons do not modify in any way
the physics of the original theory. To see this restrict the source
to be a taste singlet, $J= \widetilde J\otimes {\bf 1}$,
where $\widetilde J$ carries no taste indices.
Then
\begin{equation}
{\rm det}^{1/N_t}\Big([(D+M)\otimes {\bf 1}] + J \Big)
=
{\rm det}^{1/N_t}\Big([D+M+\widetilde J]\otimes {\bf 1}\Big)
=
{\rm det}\Big(D+M + \widetilde J\Big)
\,,
\eqn{J}
\end{equation}
so all correlation functions generated by $\widetilde J$ are those of QCD.
From the left-hand side of this equation we see that we can
obtain the correct QCD physics in the extended (rooted) theory if
we restrict ourselves to correlation functions in the taste-singlet sector.
Extended charge conservation implies that, 
if the initial state has a total taste-charge equal to zero,
then the same will be true for the final state.  
The taste-singlet sector of the extended theory is
the unitary, physical subspace for QCD with $N_f$ flavors.

Nevertheless, if we choose to calculate the correlations generated by $\widetilde J$
in the extended theory, using the middle form in \eq{J}, 
there will be contributions from taste non-singlet (and therefore Goldstone)
pions in some diagrams, which perforce must be canceled by similar contributions
in other diagrams.  For example, take $\widetilde J$ to be a Lorentz 
scalar and flavor singlet, and calculate the contribution of two-meson
intermediate states to the two-point function.  Consider
the  diagram where the ``valence'' quark lines (those coupled to $\widetilde J$)
contract in a single quark loop and there is an additional single ``sea'' quark
loop (from the determinants). This diagram
includes contributions from all $(N_t N_f)^2\!-\!1$ pions.  It is an amusing
exercise in chiral perturbation theory to check that the contributions of the 
taste non-singlet pions cancel when all diagrams are added.  Of course,
\eq{J} tells us that they must.

In the rooted staggered theory at non-zero lattice spacing, 
the cancellations of the taste non-singlet pions will no longer
be exact, because the violation of taste symmetry lifts the degeneracy of these
pions.  In particular, there is only one true (pseudo-)Goldstone boson per flavor.
The lack of cancellation means that the theory is 
not unitary in the physical, taste-singlet subspace
at non-zero $a$.
However, it is now clear that the cancellations must become exact and
unitarity must be restored  in the continuum
limit.  This is not a case of ``rather subtle cancellations which have not been
demonstrated,'' as Creutz claims.  Rather, it relies solely on the restoration
of taste symmetry in the continuum, for which there is extensive evidence, as detailed
above.
For detailed examples of how unitarity is recovered in
the continuum limit, see Refs.~\cite{PRELOVSEK,CB}.

We note here that the above framework, and in particular \eq{detJII},
can be used to derive non-singlet Ward identities of the type discussed by Creutz
in his original posting \cite{creutz06} (but not in Ref.~\cite{creutz07}).
Since the taste-singlet sector of the
extended, rooted, theory is equivalent to the original, target
theory, the existence of such Ward identities cannot invalidate
the rooted staggered approach.

A question that might have been raised
by the discussion so far is the following: Numerical simulations of pion and kaon properties
with rooted staggered fermions usually use
taste non-singlet states~\cite{davies,MILCmq,MILCfKpi,MILCsemil}.
How is this consistent with our statement above that one obtains the
correct physics with taste-{\it singlet}\/ states?
The point is that, in the continuum limit,
the extended vector symmetries show that flavor non-singlet physical states
(such as the charged pions and the kaons) have identical properties
to taste non-singlets with the same flavor quantum numbers.
So we can choose the taste that is most convenient. This is not true,
however, for any state which can mix with flavor singlets (such as the
$\pi_0$ when $m_u\ne m_d$).

\medskip
\noindent {\bf 4.}
Something that Creutz finds ``peculiar'' is the fact that
an ultraviolet limit ($a\to0$) and an infrared limit ($m\to0$) do not commute with
staggered fermions.  This kind of phenomenon is not peculiar to staggered
quarks, but is in fact common whenever the lattice theory has less symmetry than
the continuum theory.
For example, in the pure-glue theory on the lattice 
the continuum $2^{++}$ glueball splits into two states, the $E^{++}$ and
the $T_2^{++}$, because of the breaking of the continuum rotation group
to the cubic group. Typically, $m_{E^{++}}<m_{T_2^{++}}$ \cite{MORNINGSTAR},
and the mass difference is of order $a^2 \Lambda^3_{QCD}$, vanishing in the continuum limit.
Because the ultraviolet regulator has produced a mass splitting (as it does
for staggered fermions), this can lead to infrared effects.  
Define $G_E(t)$ and $G_{T_2}(t)$ to be (zero-spatial-momentum) correlators of interpolating fields
in the $E^{++}$ and $T^{++}_2$ representations, respectively, normalized so that the
lowest states are created with weight 1:
\begin{equation}\eqn{glue-corrs}
G_E(t) = \exp(-m_{E^{++}}t) + \cdots\;;\qquad\quad G_{T_2}(t)= \exp(-m_{T_2^{++}}t) + \cdots\ ,
\end{equation}
where $\cdots$ stands for the contributions of higher-mass states.
Then the long-distance (infrared) limit $t\to\infty$ does not commute with $a\to0$
for the ratio of correlators:
\begin{equation}\eqn{glue-limits}
\lim_{t\to\infty}\; \lim_{a\to0}\; \frac{G_{T_2}(t)}{G_E(t)} = 1\;;
\qquad\quad 
\lim_{a\to0}\; \lim_{t\to\infty}\; \frac{G_{T_2}(t)}{G_E(t)} = 0\;.
\end{equation}
This is the kind of phenomenon that Creutz calls ``absurd behavior for a regulator'':
A physical observable (the ratio of correlation functions) develops an infinite
derivative with respect to the cutoff at a particular point ($t=\infty$).  Of course,
once one understands what is going on, such behavior presents no problem in practice.
It is straightforward to extrapolate the masses in both
channels to the continuum limit to obtain the
physical $2^{++}$ glueball mass.
Indeed, we never work at the point $t=\infty$ on the lattice,
just as we never work at the point $m=0$.

\medskip
\noindent {\bf 5.}
The arguments in this Comment make clear that,
in order to use staggered quarks to compute physical quantities with controlled
systematic errors, it will generally be necessary to have
a detailed understanding of staggered discretization effects
and to extrapolate lattice data in lattice spacing and quark mass with chiral
forms that reflect these discretization effects.
 This is certainly true at
lattice spacings of existing ensembles ($\sim\!0.1\;$fm) where the discretization
effects are not negligible.  The needed forms are provided by
staggered chiral perturbation theory (\schpt) \cite{SCHPT}.  
With \schpt, even
rather complicated physical situations such as the scalar correlator (which is
sensitive to the anomaly) can be fit
and properly extrapolated to extract continuum physics \cite{PRELOVSEK}.  
Creutz claims that ``completely non-perturbative'' quantities such as
the baryon mass in the chiral limit  are ``particularly suspect.''
We see no fundamental difference between baryon masses
and pseudoscalar decay constants, which are also completely non-perturbative
in the chiral limit and have been accurately computed with staggered fermions \cite{MILCfKpi}.
It is true however that staggered computations in the baryon channel have been hindered
by the absence of \schpt\ for baryons, which has only recently been 
worked out \cite{BAILEY}.  With \schpt, well-controlled baryon computations with staggered
fermions are within reach.  
Of course, we do not claim that rooted staggered fermions
are the best approach to calculate all physical quantities in QCD. 
Depending on the size of the discretization errors and how
well they are understood, different fermions discretizations will produce results for
different physical quantities with systematic errors of varying sizes.
We argue, however, that nothing in Creutz's discussion raises doubts
about the {\it correctness}\/ of properly extrapolated staggered results.

\section*{Acknowledgments}
\label{sec:acknowledge}
This work was begun at the Institute for Nuclear Theory at the University of Washington,
which we thank for its hospitality. We are grateful to Martin Savage for discussions.
MG was supported in part by the Generalitat de Catalunya under the program
PIV1-2005.
CB, MG and SS were supported in part  by the US Department of Energy.
YS was supported by the Israel Science Foundation under grant no.~173/05.

\end{document}